\documentclass[aps,prb,twocolumn,showpacs]{revtex4-1}
\usepackage{graphicx}
\usepackage{amssymb}
\usepackage{amsmath}
\usepackage{appendix}

\begin{document}

\title{Spin Hall phenomenology of magnetic dynamics}

\author{Yaroslav Tserkovnyak}
\author{Scott A. Bender}
\affiliation{Department of Physics and Astronomy, University of California, Los Angeles, California 90095, USA}

\date{\today}

\begin{abstract}
We study the role of spin-orbit interactions in the coupled magnetoelectric dynamics of a ferromagnetic film coated with an electrical conductor. While the main thrust of this work is phenomenological, several popular simple models are considered microscopically in some detail, including Rashba and Dirac two-dimensional electron gases coupled to a magnetic insulator, as well as a diffusive spin Hall system. We focus on the long-wavelength magnetic dynamics that experiences current-induced torques and produces fictitious electromotive forces. Our phenomenology provides a suitable framework for analyzing experiments on current-induced magnetic dynamics and reciprocal charge pumping, including the effects of magnetoresistance and Gilbert-damping anisotropies, without a need to resort to any microscopic considerations or modeling. Finally, some remarks are made regarding the interplay of spin-orbit interactions and magnetic textures.
\end{abstract}


\maketitle

\section{Introduction}

Several new directions of spintronic research have opened and progressed rapidly in recent years. Much enthusiasm is bolstered by the opportunities to initiate and detect spin-transfer torques in magnetic metals\cite{andoPRL08,*mironNATM10,*mironNAT11,*liuPRL11sh,*liuSCI12} and insulators,\cite{kajiwaraNAT10,*sandwegPRL11,*burrowesAPL12,*hahnPRB13} which could be accomplished by variants of the spin Hall effect,\cite{haneyPRB13,*brataasNATN14} along with the reciprocal electromotive forces induced by magnetic dynamics. The spin Hall effect stands for a spin current generated by a transverse applied charge current, in the presence of spin-orbit interaction. From the perspective of angular momentum conservation, the spin Hall effect allows angular momentum to be leveraged from the stationary crystal lattice to the magnetic dynamics. A range of nonmagnetic materials from metals to topological insulators have been demonstrated to exhibit strong spin-orbit coupling, thus allowing for efficient current-induced torques.

Focusing on quasi-two-dimensional (2D) geometries, we can generally think of the underlying spin Hall phenomena as an out-of-equilibrium magnetoelectric effect that couples planar charge currents with collective magnetization dynamics. In typical practical cases, the relevant system is a bilayer heterostructure, which consists of a conducting layer with strong spin-orbit coupling and ferromagnetic layer with well-formed magnetic order. In this case, the current-induced spin torque reflects a spin angular momentum flux normal to the plane, which explains the spin Hall terminology. 

The microscopic interplay of spin-orbit interaction and magnetism at the interface translates into a macroscopic coupling between charge currents and magnetic dynamics. A general phenomenology applicable to a variety of disparate heterostructures can be inferred by considering a course-grained 2D system, which both conducts and has magnetic order as well as lacks inversion symmetry (or else the pseudovectorial magnetization would not couple linearly to the vectorial current density). In a bilayer heterostructure, the latter is naturally provided by the broken reflection symmetry with respect to its plane.

\section{General phenomenology}
\label{gp}

\begin{figure}[pb]
\includegraphics[width=\linewidth,clip=]{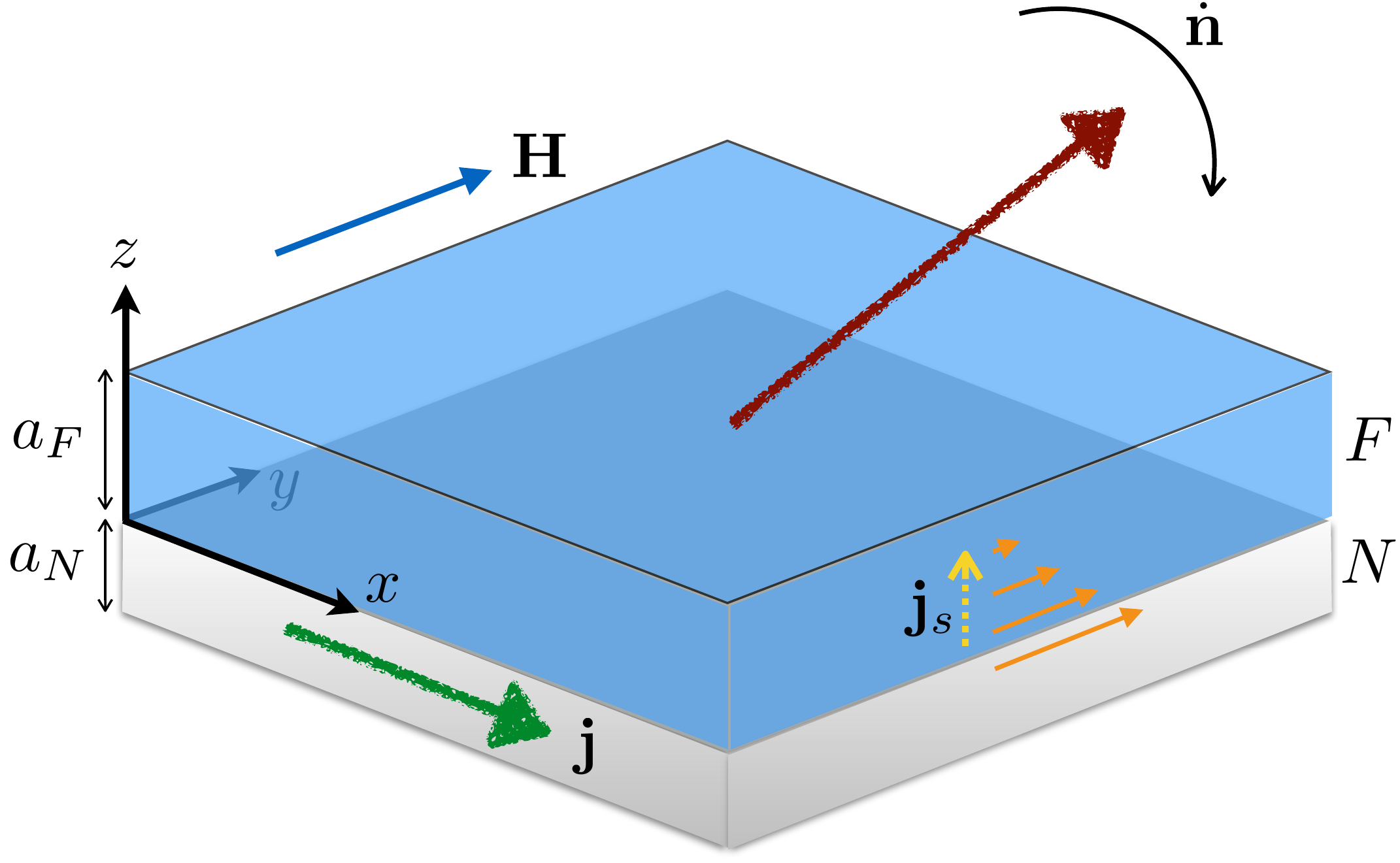}
\caption{Heterostructure consisting of a magnetic top layer and conducting underlayer. The charge current $\mathbf{j}$ induces a torque $\boldsymbol{\tau}$ acting on the magnetic dynamics, which quantifies the spin angular-momentum transfer in the $z$ direction. This can be thought of as a spin current $\mathbf{j}_s$ entering the ferromagnet at the interface. Reciprocally, magnetic dynamics $\dot{\mathbf{n}}$ induces a motive force $\boldsymbol{\epsilon}$ acting on the itinerant electrons in the conductor.}
\label{fig}
\end{figure}

Let us specifically consider a bilayer heterostructure with úone layer magnetic and one conducting, as sketched in Fig.~\ref{fig}. The nonmagnetic layer can be tailored to enhance spin-orbit coupling effects in and out of equilibrium. Phenomenologically, we have a quasi-2D system along the $xy$ plane, which will for simplicity be taken to be isotropic and mirror-symmetric in plane while breaking reflection symmetry along the $z$ axis. In other words, the structural symmetry is assumed to be that of a Rashba 2D electron gas (although microscopic details could be more complex), subject to a spontaneous time-reversal symmetry breaking due to the magnetic order. Common examples of such heterostructures include a thin transition-metal\cite{andoPRL08,*mironNATM10,*mironNAT11,*liuPRL11sh,*liuSCI12} or magnetic-insulator\cite{kajiwaraNAT10,*sandwegPRL11,*burrowesAPL12} film capped by a heavy metal, or a layer of 3D topological insulator doped on one side with magnetic impurities.\cite{chenSCI10,*wrayNATP10,*checkelskyNATP12,*fanNATM14}

The course-grained hydrodynamic variables used to describe our system are the three-component collective spin density (per unit area) $\mathbf{s}(\mathbf{r},t)=s\mathbf{n}(\mathbf{r},t)\equiv(sn_x,sn_y,sn_z)$ and the two-component 2D current density (per unit length) $\mathbf{j}(\mathbf{r},t)\equiv(j_x,j_y)$ in the $xy$ plane. Considering fully saturated magnetic state well below the Curie temperature, we treat the spin density as a directional variable, such that its magnitude $s$ is constant and orientational unit vector $\mathbf{n}$ parametrizes a smooth and slowly-varying magnetic texture. We will be interested in slow and long-wavelength agitations of the ferromagnet coupled to the electron liquid along with reciprocal motive forces. Perturbed out of equilibrium, the temporal evolution of the heterostructure is governed by the forces that couple to the charge flow and magnetic dynamics: the (planar) electric field and magnetic field, respectively.

\subsection{Decoupled dynamics}

A uniform electric-current carrying state in the isolated conducting film, subject to a constant \textit{external} vector potential $\mathbf{A}$, has the free-energy density
\begin{equation}
\mathcal{F}(\mathbf{p},\mathbf{A})=\mathcal{F}_0(\mathbf{p})-\frac{\mathbf{p}\cdot\mathbf{A}}{c}+\mathcal{O}(A^2)\,,
\end{equation}
where $\mathcal{F}_0=L\mathbf{p}^2/2$ is the free-energy density in terms of the \textit{paramagnetic} current $\mathbf{p}$ (i.e., the current defined in the absence of the vector potential $\mathbf{A}$), and $L$ is the local self-inductance of the film (including inertial and electromagnetic contributions). According to time-reversal symmetry, in equilibrium $\mathbf{p}=0$ when $\mathbf{A}=0$. The gauge invariance (which requires that the minimum of $\mathcal{F}$, as a function of $\mathbf{p}$, is independent of $\mathbf{A}$), furthermore, dictates the following form of the free energy:
\begin{equation}
\mathcal{F}=\frac{L}{2}\left(\mathbf{p}-\frac{\mathbf{A}}{cL}\right)^2\,.
\label{F}
\end{equation}
Therefore, the phenomenological expression for the full current density is
\begin{equation}
\mathbf{j}\equiv-c\delta_\mathbf{A}F=\mathbf{p}-\frac{\mathbf{A}}{cL}\,,
\label{j}
\end{equation}
with $\delta$ standing for the 2D functional derivative of the total electronic free energy $F[\mathbf{p}]=\int d^2\mathbf{r}\mathcal{F}(\mathbf{p})$. We conclude, based on Eqs.~\eqref{F} and \eqref{j}, that $\mathbf{j}=L^{-1}\delta_{\mathbf{p}}F$, which is thus the \textit{force} thermodynamically conjugate to $L\mathbf{p}$. General quasistatic equilibration\cite{landauBOOKv5} of a perturbed electron system can now be written as
\begin{equation}
L\dot{\mathbf{p}}=-\hat{\varrho}\mathbf{j}\,,
\end{equation}
or, in terms of the physical current:
\begin{equation}
L\dot{\mathbf{j}}+\hat{\varrho}\mathbf{j}=\mathbf{E}\,,
\end{equation}
where $\mathbf{E}\equiv-\partial_t\mathbf{A}/c$ is the electric field, and $\hat{\varrho}$ is identified as the resistivity tensor. This is the familiar Ohm's law, which, in steady state, reduces to
\begin{equation}
\mathbf{j}=\hat{g}\mathbf{E}\,,
\end{equation}
in terms of the conductivity tensor $\hat{g}\equiv\hat{\varrho}^{-1}$. Based on the axial symmetry around $z$, we can generally write $\hat{g}=g+g_H\mathbf{z}\times$, where $g$ is the longitudinal (i.e., dissipative) and $g_H$ Hall conductivities.

The isolated magnetic-film dynamics, on the other hand, are described by the Landau-Lifshitz-Gilbert equation:\cite{landauBOOKv9,*gilbertIEEEM04}
\begin{equation}
s(1+\alpha\mathbf{n}\times)\dot{\mathbf{n}}=\mathbf{H}^*\times\mathbf{n}\,,
\end{equation}
where $\mathbf{H}^*\equiv\delta_\mathbf{n}F[\mathbf{n}]$ is the effective magnetic field governed by the magnetic free-energy functional $F[\mathbf{n}]=\int d^2\mathbf{r}\mathcal{F}(\mathbf{n})$. The (dimensionless) Gilbert damping $\alpha$ captures the (time-reversal breaking) dissipative processes in the spin sector.

The total dissipation power in our combined, but still decoupled, system is given by
\begin{equation}
-\dot{F}=-\int d^2\mathbf{r}\left(L\dot{\mathbf{p}}\cdot\mathbf{j}+\dot{\mathbf{n}}\cdot\mathbf{H}^*\right)=\int d^2\mathbf{r}\left(\varrho j^2+\alpha s\dot{\mathbf{n}}^2\right)\,,
\label{P}
\end{equation}
where $\varrho=g/(g^2+g_H^2)$ is the longitudinal resistivity. According to the fluctuation-dissipation theorem,\cite{landauBOOKv5} finite-temperature fluctuations are thus determined by $\langle j_i(\mathbf{r},t)j_{i'}(\mathbf{r}',t')\rangle=2g k_BT\delta_{ii'}\delta(\mathbf{r}-\mathbf{r}')\delta(t-t')$ and $\langle h_i(\mathbf{r},t)h_{i'}(\mathbf{r}',t')\rangle=2\alpha sk_BT\delta_{ii'}\delta(\mathbf{r}-\mathbf{r}')\delta(t-t')$. Having mentioned this for completeness, we will not pursue thermal properties any further.

\subsection{Coupled dynamics}

Having recognized $(L\mathbf{p},\mathbf{j})$ and $(\mathbf{n},\mathbf{H}^*)$ as two pairs of thermodynamically conjugate variables, their coupled dynamics must obey Onsager reciprocity.\cite{landauBOOKv5} Charge current flowing through our heterostructure in general induces a torque $\boldsymbol{\tau}$ on the magnetic moment and, vice versa, magnetic dynamics produce a motive force $\boldsymbol{\epsilon}$ acting on the current, defined as follows:
\begin{align}
\label{llgt}
s(\dot{\mathbf{n}}+\mathbf{n}\times\hat{\alpha}\dot{\mathbf{n}})&=\mathbf{H}^*\times\mathbf{n}+\boldsymbol{\tau}\,,\\
L\dot{\mathbf{j}}+\hat{\varrho}\mathbf{j}&=\mathbf{E}+\boldsymbol{\epsilon}\,,
\label{Lj}
\end{align}
where $L\dot{\mathbf{j}}=L\dot{\mathbf{p}}+\mathbf{E}$, according to Eq.~\eqref{j}. In general, due to the spin-orbit interaction at the interface, Gilbert damping\footnote{} $\hat{\alpha}$ and resistivity tensor\cite{nakayamaPRL13,*chenPRB13} $\hat{\varrho}$ can acquire anisotropic $\mathbf{n}$-dependent contributions. Let us start by expanding the motive force, according to the assumed structural symmetries, in the Cartesian components of $\mathbf{n}$:
\begin{equation}
\boldsymbol{\epsilon}=\left[(\eta+\vartheta\mathbf{n}\times)\dot{\mathbf{n}}\right]\times\mathbf{z}\,,
\label{eps}
\end{equation}
where $\eta$ is the \textit{reactive} and $\vartheta$ the \textit{dissipative} coefficients characterizing spin-orbit interactions in our coupled system. While $\eta$ and $\vartheta$ can generally depend on $n_z^2$, we will for simplicity be focusing our attention on the limit when they are mere constants. The dimensionless parameter $\beta\equiv\vartheta/\eta$ describes their relative strengths. The Onsager reciprocity then immediately dictates the following form of the torque:
\begin{equation}
\boldsymbol{\tau}=(\eta+\vartheta\mathbf{n}\times)(\mathbf{z}\times\mathbf{j})\times\mathbf{n}\,.
\label{tau}
\end{equation}
In line with the existing nomenclature,\cite{andoPRL08,kajiwaraNAT10} we can write the dissipative coefficient as
\begin{equation}
\vartheta\equiv\frac{\hbar}{2ea_N}\tan\theta\,,
\label{vt}
\end{equation}
in terms of a length scale $a_N$, which we take to correspond to the normal-metal thickness,\footnote{} and dimensionless parameter $\theta$ identified as the effective \textit{spin Hall angle} at the interface. The coefficient $\eta$ in Eq.~\eqref{tau} parametrizes the so-called \textit{field-like torque,} which could arise, for example, as a manifestation of the interfacial \textit{Edelstein effect.}\cite{edelsteinJPCM95}

Another important effect of the nonmagnetic layer on the ferromagnet is the enhanced damping of the magnetization dynamics by spin pumping,\cite{tserkovPRL02sp,*tserkovRMP05} such that
\begin{equation}
\alpha=\alpha_0+\frac{a^{\uparrow\downarrow}}{a_F}\,.
\label{a}
\end{equation}
$\alpha_0$ is the bulk damping, which is thickness $a_F$ independent, and $a^{\uparrow\downarrow}$ parametrizes the strength of angular momentum [as well as energy, according to Eq.~\eqref{P}] loss at the interface. Spin pumping into a perfect spin reservoir corresponds to\cite{tserkovPRL02sp,*tserkovRMP05} $a^{\uparrow\downarrow}=\hbar g_r^{\uparrow\downarrow}/4\pi S$, where $g_r^{\uparrow\downarrow}$ is the (real part of the dimensionless) interfacial spin-mixing conductance per unit area and $S\equiv s/a_F$ is the 3D spin density in the ferromagnet. In reality, $a^{\uparrow\downarrow}$ depends on the spin-relaxation efficiency in the normal metal as well as the spin-orbit interaction at the interface, and may depend on $a_N$ in a nontrivial manner (see Ref.~\onlinecite{tserkovPRB02sp} for a diffusive model), so long as $a_N\lesssim\lambda_N$, where $\lambda_N$ is the spin-relaxation length in the normal metal.\footnote{} With these conventions in mind and focusing on the limit of $a_N\gg\lambda_N$ and, in the case of a metallic ferromagnet, $a_F\gg\lambda_F$, we will suppose that the coefficients $\theta$, $\beta$, and $a^{\uparrow\downarrow}$ defined above are thickness independent.\footnote{}

Unless otherwise stated, we will disregard anisotropies in $\alpha$, which may in general depend on the directions of $\mathbf{n}$ and $\dot{\mathbf{n}}$, subject to the reduced crystalline symmetries and the lack of reflection asymmetry at the interface.\footnote{} In the same spirit, with the exception of Sec.~\ref{dshs}, we will not concern ourselves much with the $\mathbf{n}$-dependent interfacial magnetoresistance/proximity effects,\cite{nakayamaPRL13} which would enter through the resistivity tensor $\hat{\varrho}(\mathbf{n})$ in Eq.~\eqref{Lj}.

We remark that while we considered a nonequilibrium magnetoelectric coupling in terms of torque $\boldsymbol{\tau}$ and force $\boldsymbol{\epsilon}$ in Eqs.~\eqref{llgt} and \eqref{Lj}, we had retained the decoupled form of the free-energy density, $\mathcal{F}(\mathbf{p})+\mathcal{F}(\mathbf{n})$. We exclude the possibility of a linear coupling of $\mathbf{p}$ to the magnetic order, since it would suggest a nonzero electric current in equilibrium.

\subsection{Current-induced instability}

Equations \eqref{llgt} and \eqref{Lj} encapsulate rich nonlinear dynamics. Of particular interest are the current-induced magnetic instabilities and switching. For a fixed current bias $\mathbf{j}$, it is convenient to multiply Eq.~\eqref{llgt} by $(1-\alpha\mathbf{n}\times)$ on the left to obtain
\begin{equation}
s(1+\alpha^2)\dot{\mathbf{n}}=\mathbf{h}\times\mathbf{n}-\alpha\mathbf{n}\times\mathbf{h}'\times\mathbf{n}\,.
\end{equation}
Here,
\begin{equation}
\mathbf{h}\equiv\mathbf{H}^*+(\eta+\vartheta\alpha)\mathbf{z}\times\mathbf{j}\,,\,\,\,\mathbf{h}'\equiv\mathbf{H}^*+(\eta-\vartheta/\alpha)\mathbf{z}\times\mathbf{j}
\label{hh}
\end{equation}
are the effective Larmor and damping fields, respectively. A magnetic instability (bifurcation) at an equilibrium fixed point may occur, for example, when either the effective field or effective damping change sign.

To illustrate this, consider a simple case, where a constant current is applied in the $x$ direction: $\mathbf{j}=j\mathbf{x}$, while an external magnetic field parametrized by  $H$ is applied along the $y$ axis: $\mathbf{H}^*=H\mathbf{y}+Kn_z\mathbf{z}$, where we also include an easy-plane magnetic anisotropy $K$. Equations \eqref{hh} then become
\begin{align}
\label{hh1}
\mathbf{h}&=\left[H+(\eta+\vartheta\alpha)j\right]\mathbf{y}+Kn_z\mathbf{z}\,,\\
\mathbf{h}'&=\left[H+(\eta-\vartheta/\alpha)j\right]\mathbf{y}+Kn_z\mathbf{z}\,.
\label{hh2}
\end{align}
In equilibrium, when $j=0$: $\mathbf{n}=-\mathbf{y}$. When $j$ is ramped up, however, this fixed point may become unstable. Let us consider two extreme limits: First, suppose the magnetoelectric coupling \eqref{tau} is purely reactive, i.e., $\vartheta=0$. The effect of the torque can thus be fully captured by a redefinition of the applied field as $H\to H+\eta j$. We thus see that when $-j$ exceeds $H/\eta$, the effective field switches sign, and the stable magnetic orientation flips from $-\mathbf{y}$ to $\mathbf{y}$.

If, on the other hand, the magnetoelectric coupling \eqref{tau} is purely dissipative, i.e., $\eta=0$, then $H\to H+\vartheta\alpha j$ according to Eq.~\eqref{hh1}, whereas $H\to H-(\vartheta/\alpha)j$ according to Eq.~\eqref{hh2}. Supposing, furthermore, that $\alpha\ll1$, as is nearly always the case, the effect of $\vartheta$ on $\mathbf{h}$ is negligible in comparison to its effect on $\mathbf{h}'$. We thus rewrite Eqs.~\eqref{hh1} and \eqref{hh2} as
\begin{equation}
\mathbf{h}\approx H\mathbf{y}+Kn_z\mathbf{z}\,,\,\,\,\mathbf{h}'=\left[H-(\vartheta/\alpha)j\right]\mathbf{y}+Kn_z\mathbf{z}\,.
\end{equation}
A simple stability analysis gives for the critical current at which $\mathbf{n}=-\mathbf{y}$ becomes unstable:
\begin{equation}
j_c=\frac{\alpha}{\vartheta}\left(H+\frac{K}{2}\right)\,.
\label{jc}
\end{equation}

In the presence of comparable reactive and dissipative torques, i.e., $\beta\sim1$ so that $\eta\sim\vartheta$, while still $\alpha\ll1$, $\mathbf{h}$ remains essentially unaffected by currents of order $j_c$ (unless $K\gtrsim H/\alpha\gg H$), so that the above dissipative magnetic instability at $j_c$ is maintained. We could thus expect Eq.~\eqref{jc} to rather generally describe the leading spin-torque instability threshold\cite{slonczewskiJMMM96} for the monodomain dynamics.

It is instructive to obtain from Eq.~\eqref{jc} the \textit{intrinsic} instability threshold for thin magnetic films, $a_F\ll a^{\uparrow\downarrow}/\alpha_0$, for which the bulk contribution, $\alpha_0$, to the damping \eqref{a} can be neglected:
\begin{equation}
j_c^{(0)}=\frac{2e}{\hbar}\frac{a^{\uparrow\downarrow}}{\tan\theta}\frac{a_N}{a_F}\left(H+\frac{K}{2}\right)\,.
\end{equation}
Writing, furthermore, $j_c^{(0)}=J_c^{(0)}a_N$, in terms of the 3D current density $J_c^{(0)}$; $a^{\uparrow\downarrow}=\hbar g_r^{\uparrow\downarrow}/4\pi S$, in terms of the \textit{effective} spin-mixing conductance $g_r^{\uparrow\downarrow}$ (including the effects of spin backflow from the normal layer,\cite{tserkovPRB02sp} in case of an imperfect spin sink); and converting effective field to physical units: $H=\omega_Ba_FS$ and $K=\omega_Ka_FS$, where $\omega_B=\gamma B$ in terms of the gyromagnetic ratio $\gamma$ and applied field $B$, $\omega_K=4\pi\gamma M_s$ with $M_s=\gamma S$, in case of only the shape anisotropy, we obtain
\begin{equation}
J_c^{(0)}=\frac{e}{2\pi}\frac{g_r^{\uparrow\downarrow}}{\tan\theta}\left(\omega_B+\frac{\omega_K}{2}\right)\,.
\end{equation}
We recall that the Kittel formula for the ferromagnetic-resonance frequency is $\omega=\sqrt{\omega_B(\omega_B+\omega_K)}$. Using quantities characteristic of the Pt$\mid$YIG compound:\cite{andoPRL08,kajiwaraNAT10} $\theta\sim0.1$, $g_r^{\uparrow\downarrow}\sim5$~nm$^{-2}$, $\omega_K\sim4\times10^{10}$~s$^{-1}$, we would get for the intrinsic instability threshold (in the absence of an applied field $B$): $J_c^{(0)}\sim3\times10^{10}$~A$\cdot$m$^{-2}$. (Threshold currents at this order were also evaluated in Ref.~\onlinecite{zhouPRB13}.) In the opposite limit of thick magnetic films, $a_F\gg a^{\uparrow\downarrow}/\alpha_0$ ($\sim1/2$~$\mu$m for YIG, using $\alpha_0\sim10^{-4}$), the bulk Gilbert damping dominates magnetic dissipation, and
\begin{equation}
J_c\approx\frac{\alpha_0a_F}{a^{\uparrow\downarrow}}J_c^{(0)}=\frac{2e}{\hbar}\frac{\alpha_0a_FS}{\tan\theta}\left(\omega_B+\frac{\omega_K}{2}\right)
\end{equation}
increases linearly with $a_F$ beyond the intrinsic threshold.

\section{simple models}
\label{sm}

Equations \eqref{llgt}-\eqref{a} provide a general phenomenological framework for exploring the coupled magnetoelectric dynamics in thin-film magnetic heterostructures, which we verify by considering several simple microscopic models in the following.

\subsection{Rashba Hamiltonian}
\label{RH}

One of the simplest models engendering the phenomenology of interest is based on a 2D electron gas at a reflection-asymmetric interface, which, at low energies, is described by the (single-particle) Rashba Hamiltonian:
\begin{equation}
\hat{H}_R=\frac{p^2}{2m}+v\mathbf{p}\cdot\mathbf{z}\times\hat{\boldsymbol{\sigma}}\,.
\label{HR}
\end{equation}
Velocity $v$ here parametrizes the spin-orbit interaction strength due to structural asymmetry; $\hat{\boldsymbol{\sigma}}$ is a vector of Pauli matrices. When the first (nonrelativistic) term in Hamiltonian \eqref{HR} dominates over the second (relativistic) term (i.e., $v\ll v_F$, the Fermi velocity), we can treat $v$ perturbatively.

To zeroth order in $v$, the velocity operator is $\partial_\mathbf{p}\hat{H}_R=\mathbf{p}/m$, such that the current density is $\mathbf{j}=-en\langle\mathbf{p}\rangle/m$, in terms of the particle-number density $n=k_F^2/2\pi=m^2v_F^2/2\pi\hbar^2$ and the positron charge $e>0$. On the other hand, to first order in $v$, Eq.~\eqref{HR} results in the steady-state spin density
\begin{equation}
\boldsymbol{\rho}=\frac{mv}{2\pi\hbar}\mathbf{z}\times\langle\mathbf{p}\rangle=-\frac{m^2v}{2\pi\hbar en}\mathbf{z}\times\mathbf{j}\,,
\label{rho}
\end{equation}
recalling that the 2D density of states (which defines the spin susceptibility) is given by $m/2\pi\hbar^2$. Equation \eqref{rho} reflects the Edelstein effect.\cite{edelsteinJPCM95}

Exchange coupling this Rashba 2DEG to an adjacent ferromagnet according to the local Hamiltonian
\begin{equation}
H'=-\int d^2\mathbf{r}\left[J\left(n_x\rho_x+n_y\rho_y\right)+J_\perp n_z\rho_z\right]\,,
\label{Hp}
\end{equation}
where $J$ and $J_\perp$ are respectively the in-plane and out-of-plane exchange constants, we get for the torque:
\begin{equation}
\boldsymbol{\tau}=\delta_\mathbf{n}H'\times\mathbf{n}=-\left[J(\rho_x\mathbf{x}+\rho_y\mathbf{y})+J_\perp\rho_z\mathbf{z}\right]\times\mathbf{n}\,.
\label{tauR}
\end{equation}
Evaluating this torque to leading (i.e., first) order in the exchange, we need to find $\boldsymbol{\rho}$ to zeroth order, which is given by Eq.~\eqref{rho}. We thus have:
\begin{equation}
\boldsymbol{\tau}=\eta(\mathbf{z}\times\mathbf{j})\times\mathbf{n}\,,
\end{equation}
where
\begin{equation}
\eta=\frac{m^2vJ}{2\pi\hbar en}=\frac{\hbar}{e}\frac{vJ}{v_F^2}\,.
\label{eR}
\end{equation}
The dissipative (i.e., spin Hall) coefficient $\vartheta$ vanishes in this model at this level of approximation. We should, however, expect $\vartheta$ to arise at quadratic order in $J$ [whereas at first order in $J$, it must vanish for arbitrarily large $v$, since, in the absence of magnetism, Eq.~\eqref{rho} here describes the general form of spin response to dc current].

\subsection{Dirac Hamiltonian}
\label{DH}

In the opposite extreme, the spin-orbit interaction in Eq.~\eqref{HR} dominates over the nonrelativistic piece, which formally corresponds to sending $m\to\infty$. The corresponding 2D Dirac Hamiltonian
\begin{equation}
\hat{H}_D=v\mathbf{p}\cdot\mathbf{z}\times\hat{\boldsymbol{\sigma}}
\label{HD}
\end{equation}
arises physically on the surfaces of strong 3D topological insulators.\cite{pankratovCHA91,*hasanRMP10,*qiRMP11}

Exchange coupling electrons to a magnetic order $\mathbf{n}$, according to Eq.~\eqref{Hp}, gives the single-particle Hamiltonian
\begin{equation}
\hat{H}'=-\frac{\hbar}{2}\left[J\left(n_x\hat{\sigma}_x+n_y\hat{\sigma}_y\right)+J_\perp n_z\hat{\sigma}_z\right]\,,
\label{HDp}
\end{equation}
which can be combined with Eq.~\eqref{HD} as follows:
\begin{equation}
\hat{H}_D+\hat{H}'=v(\mathbf{p}-\mathbf{A}^\ast)\cdot\mathbf{z}\times\hat{\boldsymbol{\sigma}}-m^\ast\hat{\sigma}_z\,.
\end{equation}
Here,
\begin{equation}
\mathbf{A}^\ast\equiv\frac{\hbar J}{2v}\mathbf{z}\times\mathbf{n}~~~{\rm and}~~~m^\ast\equiv\frac{\hbar J_\perp}{2}n_z
\end{equation}
are fictitious vector potential and mass. The corresponding electromotive force (recalling that the electron charge is $-e$) is
\begin{equation}
\boldsymbol{\epsilon}=\frac{\partial_t\mathbf{A}^\ast}{e}=-\frac{\hbar J}{2ev}\dot{\mathbf{n}}\times\mathbf{z}\,,
\end{equation}
such that, according to Eq.~\eqref{eps},
\begin{equation}
\eta=-\frac{\hbar J}{2ev}\,,
\label{eD}
\end{equation}
which is of opposite sign to Eq.~\eqref{eR}. Note that unlike the latter result, Eq.~\eqref{eD} is derived nonperturbatively.

The reciprocal torque \eqref{tau} with this $\eta$ gives:
\begin{equation}
\boldsymbol{\tau}=\eta(\mathbf{z}\times\mathbf{j})\times\mathbf{n}\,.
\label{tauD}
\end{equation}
Using the helical identity between the current and spin densities,
\begin{equation}
\mathbf{j}=-\frac{2ev}{\hbar}\mathbf{z}\times\boldsymbol{\rho}\,,
\end{equation}
according to the velocity operator  $\partial_\mathbf{p}\hat{H}_D=v\mathbf{z}\times\hat{\boldsymbol{\sigma}}$, we recognize in Eq.~\eqref{tauD} the torque \eqref{tauR} due to the planar exchange $J$. The above relations mimic the structure of the preceding Rashba model. For a vanishing chemical potential, the mass term opens a gap, in which case the long-wavelength conductivity tensor is given by the half-quantized Hall response:\cite{redlichPRL84,*jackiwPRD84} $\hat{g}=-{\rm sgn}(m^\ast)(e^2/4\pi\hbar)\mathbf{z}\times$. In addition to the in-plane spin density $\mathbf{z}\times\boldsymbol{\rho}\times\mathbf{z}$ entering Eq.~\eqref{tauD}, the out-of-plane component $\rho_z$ should also exert a torque $\propto J_\perp$, according to the exchange coupling \eqref{tauR}. At the leading order, the latter contributes to the out-of-plane magnetic anisotropy $K$, which is absorbed by the magnetic free-energy density $\mathcal{F}(\mathbf{n})$.\cite{tserkovPRL12} At a finite doping, the $J_\perp$ interaction could in general be also expected to give rise to a dissipative coupling $\vartheta$.

\subsection{Diffusive spin Hall system}
\label{dshs}

The previous two models naturally produced the reactive coupling $\eta$ between planar charge current and magnetic dynamics. Here, we recap a diffusive spin Hall model\cite{mosendzPRL10,nakayamaPRL13} that results in both $\eta$ and $\vartheta$, which is based on a film of a featureless isotropic normal-metal conductor in contact with ferromagnetic insulator. If electrons diffuse through the conductor with weak spin relaxation, we can develop a hydrodynamic description based on continuity relations both for spin and charge densities. We first construct bulk diffusion equations and then impose spin-charge boundary conditions, which allows us to solve for spin-charge fluxes in the normal metal and torque on the ferromagnetic insulator.

The relevant hydrodynamic quantities in the normal-metal bulk are 3D charge and spin densities, $\rho(\mathbf{r},t)$ and $\boldsymbol{\rho}(\mathbf{r},t)$, respectively. The associated thermodynamic conjugates are the electrochemical potential, $\mu\equiv-e\delta_\rho F$, and spin accumulation, $\boldsymbol{\mu}\equiv\hbar\delta_{\boldsymbol{\rho}} F$, where $F[\rho,\boldsymbol{\rho}]$ is the free-energy functional of the normal metal. Supposing only a weak violation of spin conservation (due to magnetic or spin-orbit impurities), we phenomenologically write spin-charge continuity relations as
\begin{equation}
\partial_t\rho=-\partial_\imath J_\imath\,,\,\,\,\partial_t\rho_\jmath=-\partial_\imath J_{\imath\jmath}-\Gamma\mu_\jmath\,,
\end{equation}
where $\imath$ and $\jmath$ label Cartesian components of real and spin spaces, respectively, and the summation over the repeated index $\imath$ is implied. $\Gamma=\hbar\mathcal{N}/2\tau_s$, in terms of the (per spin) Fermi-level density of states $\mathcal{N}$ and spin-relaxation time $\tau_s$. $J_\imath$ are the components of the 3D vectorial charge-current density and $J_{\imath\jmath}$ of the tensorial spin-current density, which can be expanded in terms of the thermodynamic forces governed by $\mu$ and $\boldsymbol{\mu}$:
\begin{align}
\label{ji}
J_\imath&=\frac{\sigma}{e}\partial_\imath\mu-\frac{\sigma'}{2e}\epsilon_{\imath\jmath k}\partial_\jmath\mu_k\,,\\
\frac{2e}{\hbar}J_{\imath\jmath}&=-\frac{\sigma_+}{2e}\partial_\imath\mu_\jmath-\frac{\sigma_-}{2e}\partial_\jmath\mu_\imath-\frac{\sigma'}{e}\epsilon_{\imath\jmath k}\partial_k\mu\,,
\label{jij}
\end{align}
where $\sigma$ is the (isotropic) electrical conductivity and $\sigma'$ the spin Hall conductivity of the normal-metal bulk. The last terms of Eqs.~\eqref{ji} and \eqref{jij} are governed by the same coefficient $\sigma'$ due to the Onsager reciprocity. The bulk spin Hall angle $\theta'$ is conventionally defined by
\begin{equation}
\tan\theta'\equiv\frac{\sigma'}{\sigma}\,.
\end{equation}

Bulk diffusion equations \eqref{ji}, \eqref{jij} are complemented by the boundary conditions
\begin{equation}
J_z=0\,\,\,{\rm at}~z=-a_N,0
\end{equation}
for the charge current, where $z=-a_N$ corresponds to the normal-metal interface with vacuum and $z=0$ to the interface with the ferromagnet, and\cite{tserkovPRL02sp}
\begin{equation}
\mathbf{J}_z=\frac{1}{4\pi}\left\{
\begin{array}{cl}
0&{\rm at}~z=-a_N\\
\left(g_i^{\uparrow\downarrow}+g_r^{\uparrow\downarrow}\mathbf{n}\times\right)\tilde{\boldsymbol{\mu}}\times\mathbf{n}&{\rm at}~z=0
\end{array}\right.\,,
\end{equation}
for the spin current, with $\mathbf{J}_z$ standing for $J_{z\jmath}$. Here, $\tilde{\boldsymbol{\mu}}\equiv\boldsymbol{\mu}-\hbar\mathbf{n}\times\dot{\mathbf{n}}$ captures contributions from the spin-transfer torque and spin pumping, respectively.

Having established the general structure of the coupled spin and charge diffusion, let us calculate the steady-state charge-current density $\mathbf{j}$ driven by a simultaneous application of a uniform electric field in the $xy$ plane, $\boldsymbol{\nabla}\mu\to e\mathbf{E}$, and magnetic dynamics, $\dot{\mathbf{n}}$:
\begin{equation}
\mathbf{J}=\sigma\mathbf{E}-\frac{\sigma'}{2e}\boldsymbol{\nabla}\times\boldsymbol{\mu}\,.
\label{jmu}
\end{equation}
The spin accumulation $\boldsymbol{\mu}$ is found by solving
\begin{equation}
\left(\frac{\sigma_+}{\sigma}+\frac{\sigma_-}{\sigma}\delta_{zj}\right)\partial_z^2\mu_j=\frac{\mu_j}{l_s^2}\,,
\label{sd}
\end{equation}
where $l_s\equiv\sqrt{\hbar \sigma/4e^2\Gamma}$ is the spin-diffusion length. Using Drude formula for the conductivity $\sigma$, we get the familiar $l_s=l/\sqrt{3\epsilon}$, where $l$ is the scattering mean free path and $\epsilon\equiv\tau/\tau_s\ll1$ is the spin-flip probability per scattering ($\tau$ is the transport mean free time). The boundary conditions are
\begin{align}
\sigma'\mathbf{z}\times\mathbf{E}-\frac{\sigma_+}{2e}\partial_z\boldsymbol{\mu}-\frac{\sigma_-}{2e}\boldsymbol{\nabla}\mu_z&\nonumber\\
&\hspace{-4cm}=\frac{e}{h}\left\{
\begin{array}{cl}
0&{\rm at}~z=-a_N\\
\left(g_i^{\uparrow\downarrow}+g_r^{\uparrow\downarrow}\mathbf{n}\times\right)\tilde{\boldsymbol{\mu}}\times\mathbf{n}&{\rm at}~z=0
\end{array}\right.\,,
\label{bc}
\end{align}
where $h=2\pi\hbar$ is the Planck's constant.

In the limit of vanishing spin-orbit coupling, $\sigma_+\to \sigma$, $\sigma_-\to0$, and $\theta'\to0$. For small but finite spin-orbit interaction, we may expect $(\sigma_+-\sigma)\sim \sigma_-\sim\mathcal{O}(\theta'^2)$. In the following, we will neglect these quadratic terms and approximate $\tan\theta'\approx\theta'\ll1$, in the spirit of the present construction.

In the limit of $l_s\ll a_N$, the spin accumulation decays exponentially away from the interface as $\boldsymbol{\mu}(z)=\boldsymbol{\mu}_0e^{z/l_s}$, where
\begin{align}
\boldsymbol{\mu}_0=(\xi_i+\xi\mathbf{n}\times)\left[\hbar\dot{\mathbf{n}}-2el_s\theta'(\mathbf{z}\times\mathbf{E})\times\mathbf{n}\right]+2el_s\theta'\mathbf{z}\times\mathbf{E}\,.
\end{align}
Here, $\xi\equiv\chi(1+\zeta+\zeta_i^2)$ and $\xi_i\equiv\chi\zeta\zeta_i$, in terms of $\zeta\equiv\sigma/g_Qg^{\uparrow\downarrow}_rl_s$, $\zeta_i\equiv g^{\uparrow\downarrow}_i/g^{\uparrow\downarrow}_r$, $\chi^{-1}\equiv(1+\zeta)^2+\zeta_i^2$, and the quantum of conductance $g_Q\equiv2e^2/h$. The spin accumulation $\boldsymbol{\mu}_0$ consists of the decoupled spin-pumping and spin Hall contributions. Integrating the resultant charge-current density \eqref{jmu} over the normal-layer thickness $a_N$, we finally get for the 2D current density in the film:
\begin{equation}
\mathbf{j}=\sigma\left(a_N\mathbf{E}-\frac{\theta'}{2e}\mathbf{z}\times\boldsymbol{\mu}_0\right)=\hat{g}\left\{\mathbf{E}+\left[(\eta+\vartheta\mathbf{n}\times)\dot{\mathbf{n}}\right]\times\mathbf{z}\right\}\,,
\label{jm}
\end{equation}
where
\begin{equation}
\frac{\hat{g}}{\sigma}=\tilde{a}_N+l_s\theta'^2\left\{\xi_in_z(\mathbf{z}\times)-\xi[n_z^2+(\mathbf{z}\times\mathbf{n}\times\mathbf{z})\mathbf{n}\cdot]\right\}
\end{equation}
is the anisotropic 2D conductivity tensor ($\tilde{a}_N\equiv a_N+l_s\theta'^2\approx a_N$), which is referred in the literature to as the spin Hall magnetoconductance,\cite{nakayamaPRL13} and
\begin{equation}
\eta\approx\frac{\hbar}{2ea_N}\theta'\xi_i\,,\,\,\,\vartheta\approx\frac{\hbar}{2ea_N}\theta'\xi\,,
\end{equation}
neglecting corrections that are cubic in $\theta'$. If $\zeta_i\ll1$, which is typically the case,\cite{brataasPRP06} we have $\vartheta\gg\eta$. It could be noted that restoring $\sigma_-\sim\mathcal{O}(\theta'^2)$ in Eqs.~\eqref{sd} and \eqref{bc} would affect $\hat{g}$ only at order $\mathcal{O}(\theta'^3)$.

The above spin accumulation can also be used to calculate the spin-current density injected into the ferromagnet at $z=0$:
\begin{align}
\mathbf{J}_z&=\frac{\hbar\sigma}{2e}\left(\theta'\mathbf{z}\times\mathbf{E}-\frac{\boldsymbol{\mu}_0}{2el_s}\right)\nonumber\\
&\approx-s\mathbf{n}\times\hat{\alpha}\dot{\mathbf{n}}+(\eta+\vartheta\mathbf{n}\times)(\mathbf{z}\times\mathbf{j})\times\mathbf{n}\,,
\label{Jz}
\end{align}
where
\begin{equation}
\hat{\alpha}=\frac{\hbar^2\sigma}{4e^2l_ss}(\xi-\xi_i\mathbf{n}\times)\,,
\label{ah}
\end{equation}
and we dropped terms that are cubic in $\theta'$, as before. The corresponding magnetic equation of motion $s\dot{\mathbf{n}}=\mathbf{H}^*\times\mathbf{n}+\mathbf{J}_z$ reproduces Eq.~\eqref{Lj}, with the current-driven torque of the form \eqref{tau} that is Onsager reciprocal to the motive force in Eq.~\eqref{jm}. Writing the Gilbert damping $\propto\xi$ in Eq.~\eqref{ah} as $a^{\uparrow\downarrow}/a_F$ identifies the interfacial damping enhancement in Eq.~\eqref{a}. In the formal limit $\sigma\to\infty$ (while keeping all other parameters, including $l_s$, fixed), which reproduces the perfect spin sink, this gives $a^{\uparrow\downarrow}=\hbar g_r^{\uparrow\downarrow}/4\pi S$. In the general case, $\xi$ also captures the spin backflow from the normal layer.\cite{tserkovPRB02sp} An anisotropic contribution to the Gilbert damping would be produced at the cubic order in $\theta'$, had we not made any approximations in Eq.~\eqref{Jz}.

\section{magnetic textures}

For completeness, we also provide some rudimentary remarks regarding the effect of directional magnetic inhomogeneities, such as those associated with, for example, magnetic domain walls.\cite{emoriNATM13,*ryuNATN13} Expanding the 2D magnetic free-energy density to second order in spatial derivatives, we have for a film with broken reflection symmetry in the $xy$ plane (see Sec.~\ref{gp} for a detailed description of the structure shown in Fig.~\ref{fig}):\cite{bogdanovJMMM94}
\begin{equation}
\mathcal{F}(\mathbf{n})=\mathbf{n}\cdot\mathbf{H}+\frac{K}{2}n_z^2+\Gamma\left(n_z\partial_in_i-n_i\partial_in_z\right)+\frac{A}{2}(\partial_i\mathbf{n})^2\,,
\label{Fdm}
\end{equation}
where summation over Cartesian coordinates $i=x,y$ is implied and the dot products are in the 3D spin space. $\Gamma$ here parametrizes the strength of the Dzyaloshinski-Moriya (DM) interaction and $A$ is the magnetic exchange stiffness. A nonzero $\Gamma$ requires macroscopic breaking of the reflection symmetry as well as a microscopic spin-orbit interaction that breaks the spin-space isotropicity. Equation~\eqref{Fdm} can be rewritten in a more compact form as
$\partial_x\mathbf{n}(\mathbf{y}\times\mathbf{n})-\partial_y\mathbf{n}(\mathbf{x}\times\mathbf{n})=-n_x\partial_xn_z+n_z\partial_xn_x-n_y\partial_yn_z+n_z\partial_yn_y$
\begin{equation}
\mathcal{F}(\mathbf{n})=\mathbf{n}\cdot\mathbf{H}+\frac{\tilde{K}}{2}n_z^2+\frac{A}{2}(D_i\mathbf{n})^2\,,
\end{equation}
where
\begin{equation}
D_i\equiv\partial_i+Q\left(\mathbf{z}\times\mathbf{e}_i\right)\times
\label{Di}
\end{equation}
is the so-called chiral derivative,\cite{kimPRL13} $Q\equiv\Gamma/A$, and $\tilde{K}\equiv K-\Gamma^2/A$. $Q$ is the wave number of the magnetic spiral that minimizes the texture-dependent part of the free energy.

The DM interaction of the form \eqref{Fdm} arises naturally from the Rashba Hamiltonian \eqref{HR}. In a minimal model,\cite{kimPRL13} where electrons with the single-particle Hamiltonian \eqref{HR} magnetically order due to their spin-independent (e.g., Coulombic) interaction, the spin-orbit term $\propto v$ can be gauged out at the first order in $v$ by a position-dependent rotation in spin space. To see this, we first rewrite Eq.~\eqref{HR} as
\begin{equation}
\hat{H}_R=\frac{p^2}{2m}+v\mathbf{p}\cdot\mathbf{z}\times\hat{\boldsymbol{\sigma}}=\frac{\left(\mathbf{p}+mv\mathbf{z}\times\hat{\boldsymbol{\sigma}}\right)^2}{2m}-mv^2\,.
\label{HR2}
\end{equation}
It then immediately follows that
\begin{equation}
\hat{U}^\dagger\hat{H}_R\hat{U}=\frac{p^2}{2m}+\mathcal{O}(v^2)\,,\,\,\,{\rm where}\,\,\,\hat{U}=e^{-iQ_R\mathbf{r}\cdot\mathbf{z}\times\hat{\boldsymbol{\sigma}}/2}\,,
\label{HRU}
\end{equation}
defining
\begin{equation}
Q_R\equiv\frac{2mv}{\hbar}\,.
\end{equation}
$\hat{U}$ is the operator of spin rotation around axis $\mathbf{r}\times\mathbf{z}$ by angle $rQ_R$ (recalling that $\mathbf{r}\in xy$ plane), such that the electron spin precesses by angle $2\pi$ over distance $l_{\rm so}\equiv2\pi/Q_R=h/2mv$ (the spin-precession length). Since the transformed Hamiltonian \eqref{HRU} would describe magnetic order that is spin isotropic, the corresponding free energy is given simply by $(A/2)(\partial_i\mathbf{n})^2$ (neglecting external and dipolar fields). In the original frame of reference with Rashba Hamiltonian \eqref{HR2}, the free-energy density is then given by $\mathcal{F}(\mathbf{n})=(A/2)(\partial_i\tilde{\mathbf{n}})^2$, where $\mathbf{n}=\hat{R}\tilde{\mathbf{n}}$ and $\hat{R}(\mathbf{r})$ is the natural SO(3) representation of $\hat{U}(\mathbf{r})$. Differentiating $\partial_i\tilde{\mathbf{n}}=\hat{R}^T(\partial_i+\hat{R}\partial_i\hat{R}^T)\mathbf{n}$, we finally obtain $\mathcal{F}(\mathbf{n})=(A/2)(D_i\mathbf{n})^2$, where
\begin{equation}
D_i=\partial_i+\hat{R}\partial_i\hat{R}^T=\partial_i+Q_R\left(\mathbf{z}\times\mathbf{e}_i\right)\times
\end{equation}
indeed reproduces Eq.~\eqref{Di} with $Q\to Q_R$. In Ref.~\onlinecite{tserkovPRL12}, the free-energy density \eqref{Fdm} was also obtained for the Dirac model of Sec.~\eqref{DH}, with the result:
\begin{equation}
\Gamma_D\sim-\frac{\hbar}{8\pi v}JJ_\perp\,.
\end{equation}

As was pointed out in Ref.~\onlinecite{kimPRL13}, the chiral derivative \eqref{Di} is also expected to govern the nonequilibrium magnetic-texture properties such as the current-driven torque $\boldsymbol{\tau}$ and the spin-motive force $\boldsymbol{\epsilon}$. This can either be derived microscopically or understood on purely phenomenological symmetry-based grounds. For example, the hydrodynamic (advective) spin-transfer torque (along with its Onsager-reciprocal motive force) \cite{tserkovJMMM08}
\begin{equation}
\boldsymbol{\tau}\propto(\mathbf{j}\cdot\boldsymbol{\nabla})\mathbf{n}\,,
\end{equation}
which arises due to spin-current continuity in a model without any spin-orbit interactions and frozen magnetic impurities, would be modified by replacing $\boldsymbol{\nabla}\to\mathbf{D}$ in the perturbative treatment of the above Rashba model. However, while this simplifies a phenomenological construction of various terms, in general, there is no fundamental reason why the same $Q$ should define the chiral derivatives entering in different physical properties (such as free energy and spin torque).

\section{Conclusions}

In summary, we have developed a phenomenology for slow long-wavelength dynamics of conducting quasi-2D magnetic films and heterostructures, subject to structural symmetries and Onsager reciprocity. The formalism could address both small- and large-amplitude magnetic precession (assuming it is slow on the characteristic electronic time scales), including, for example, magnetic switching and domain-wall or skyrmion motion. Owing to the versatility of available heterostructures, including those based on magnetic and topological insulators, we have focused our discussion on the case of a ferromagnetic/nonmagnetic bilayer, which serves two purposes: It naturally has a broken inversion symmetry, and the spin-orbit and magnetic properties could be separately optimized and tuned in one of the two layers.

In the case when the spin-relaxation length in the normal layer is short compared to its thickness, we can associate the interplay between spin-orbit and exchange interactions to a narrow region in the vicinity of the interface, for which we define the kinetic coefficients such as the interfacially enhanced Gilbert damping parametrized by $a^{\uparrow\downarrow}$ and the spin Hall angle parametrized by $\vartheta$. Such (separately measurable) phenomenological coefficients, which enter in our theory, must thus be viewed as joint properties of both of the bilayer materials as well as structure and quality of the interface.

We demonstrate the emergence of our phenomenology out of three microscopic models, based on Rashba, Dirac, and diffusive normal-metal films, all in contact with a magnetic insulator. In addition to Onsager-reciprocal spin-transfer torques and electromotive forces, our phenomenology also accommodates arbitrary Gilbert-damping and (magneto)resistance anisotropies, which are dictated by the same structural symmetries and may microscopically depend on the same exchange and spin-orbit ingredients as the reciprocal magnetoelectric coupling effects.

\acknowledgments

We acknowledge stimulating discussions with G.~E.~W. Bauer, S.~T.~B. Goennenwein, and D.~C. Ralph. This work was supported in part by FAME (an SRC STARnet center sponsored by MARCO and DARPA), the NSF under Grant No. DMR-0840965, and by the Kavli Institute for Theoretical Physics through Grant No. NSF PHY11-25915.

%\bibliography{ref}

%merlin.mbs apsrev4-1.bst 2010-07-25 4.21a (PWD, AO, DPC) hacked
%Control: key (0)
%Control: author (8) initials jnrlst
%Control: editor formatted (1) identically to author
%Control: production of article title (-1) disabled
%Control: page (0) single
%Control: year (1) truncated
%Control: production of eprint (0) enabled
%

\end{document}